# Surface Plasmonic Lattice Solitons

Yao Kou, Fangwei Ye,* and Xianfeng Chen

*Department of Physics, The State Key Laboratory on Fiber Optic Local Area Communication Networks and Advanced Optical Communication Systems, Shanghai Jiao Tong University, Shanghai 200240, China*
*Corresponding author: fangweiye@sjtu.edu.cn*

*We reveal the existence of the surface plasmonic lattice solitons (surface PLSs) at the boundary of a semi-infinite metallic-dielectric periodic nano-structure. We find that the truncation of the periodic structure imposes a threshold power for the existence of surface PLSs, and enhances significantly the modal localization. The propagation and excitation of surface PLSs as well as their potential application in the all-optical subwavelength switching are also demonstrated.*

Light interaction with photonic periodic lattice has been studied extensively due to its fundamental and practical importance [1]. The study of such interaction in the presence of nonlinear optical response has also drawn great attentions as the nonlinearity affords light-controlled tunability [2, 3]. Discrete solitons [4, 5], or more generally lattice solitons, are one of the most exciting outcomes of such nonlinear interaction, and they were investigated in perfect lattice and various inhomogeneous lattice structures. Especially worthy to be mentioned is the so called surface lattice solitons occurring at the boundary of a truncated lattice [6-8], which differ profoundly from their counterpart in homogenous lattices. However, so far surface lattice solitons are all studied at the edge of *dielectric* waveguides.

There is an increasing interest of pushing the study of nonlinear light-lattice interaction into the subwavelength region, where the conventional dielectric lattice is replaced by a plasmonic one [9-12]. Plasmonic lattice is composed of alternative nanoscale metallic and dielectric materials. Tunneling of surface plasmonic polaritons (SPPs) between adjacent metallic components in plasmonic lattices might be inhibited by the nonlinearity of the dielectric medium, leading to the formation of so-called plasmonic lattice solitons (PLSs) [10]. In this Letter, we truncate the plasmonic lattice and consider the influence of the resulting lattice boundary on PLSs. We reveal the existence of a new type of nonlinear surface states, i.e., surface plasmonic lattice solitons (surface PLSs). The properties of surface PLSs are crucially determined by the boundary effect. This include the occurrence of a threshold power for their formation and the double enhanced modal localization in comparisons with their PLS counterparts. The propagation and excitation of surface PLSs are studied in detail, and their potential application in all-optical switching at the deep-subwavelength scale is also presented.

We thus consider the wave propagation along the interface between a uniform media ($x<0$) and a one-dimensional nonlinear plasmonic lattice ($x>0$). Without loss of generality, unless otherwise stated, the width of metal and Kerr-type nonlinear dielectric layers are fixed as $t_m$=60 nm and $t_d$=100 nm, respectively. The refractive index of the nonlinear layers is assumed to be intensity-dependent as $n_{NL} = \sqrt{\varepsilon_{NL}} = 1.5 + n_2 I$, where $n_2 = \pm 1.8 \times 10^{-17} m^2/W$ is the Kerr coefficient and the positive/negative sign represents self-focusing/self-defocusing nonlinearity, respectively. Intensity $I = \frac{1}{2}\varepsilon_0 c n_0 |E|^2$. The complex permittivity of the metal (silver) $\varepsilon_m$ is taken from Ref. [13], which is $\varepsilon_m$ = -129+3.28$i$ for the wavelength of $\lambda$=1550 nm. Finally, the uniform region is assumed to be linear with a permittivity $\varepsilon_d$ =2.25. The results do not change visibly if the intensity-dependent permittivity is also included in this region.

To describe nonlinear stationary modes localized at the surface of the lattice, we consider TM waves ($E_y$=$H_x$=$H_z$=0) with the stationary form

$$E_x(x,z,t) = [u_x(x)\hat{x}]e^{i(\beta z - \omega t)} \quad (1)$$

$$H_y(x,z,t) = [v_y(x)\hat{y}]e^{i(\beta z - \omega t)} \quad (2)$$

where $u_x$ and $v_y$ represent modal profile independent of $z$ ($z$ is the propagation direction), $\beta$ being soliton propagation constant. Substituting Eq. (1) and (2) into Maxwell's equations yields the following equations [9]

$$\frac{k_0 \varepsilon}{z_0} u_x = \beta v_y \quad (3)$$

$$\frac{z_0}{k_0}\left[\frac{d}{dx}(\frac{1}{\varepsilon}\frac{d}{dx}) + k_0^2\right]v_y = \beta u_x \quad (4)$$

where $z_0 = \sqrt{\mu_0/\varepsilon_0}$ is the vacuum impedance and $\varepsilon$ denotes the material permittivity as a function of $x$. $k_0 = 2\pi/\lambda$. The soliton solutions were numerically found by a self-consistent method [14]. For simplicity, the solutions given below are found without taking into account the metallic loss. However, we confirmed that inclusion of loss does not change significantly the results.

Surface PLSs are found to be centered their peak amplitude in different nonlinear dielectric layers. Figures 1 shows representative profiles whose peak amplitudes reside at the first or the second waveguide (denoted as

m=1 and m=2 modes, respectively) under focusing/defocusing Kerr-type nonlinearities. It can be seen that the solitons' amplitudes decay exponentially fast from their peak-amplitude-resided waveguide into both the uniform and the lattice regions. However, due to the boundary effect, the decay speed of amplitude in the uniform region is faster than that in the lattice region. Thus surface PLSs feature asymmetric profiles. Their asymmetry gets quickly insignificant when their peak amplitude shifts to the deep lattice region. In fact, for nonlinear index change of $\Delta n$=0.05, surface PLSs of m=4 are visually symmetric already. Note that the appearance of staggered (unstaggered) surface PLSs in the focusing (defocusing) nonlinearity is a consequence of the inverted diffraction relations that is unique to the plasmonic lattice systems [9, 10].

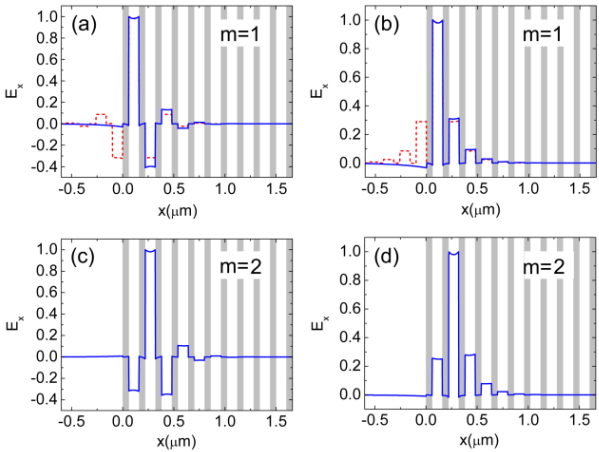

Fig. 1. Normalized electric field ($E_x$) profiles of surface PLSs for the nonlinear index change of (a), (c) $\Delta n$=0.05; (b), (d) $\Delta n$ = -0.05. The gray regions stand for metallic layers, while the white regions for dielectric domains. The red (dashed) lines in (a) and (b) represent the electric field ($E_x$) profiles of PLSs in the corresponding homogenous plasmonic lattices.

The characteristic power of surface solitons, defined as $P = (1/2)\int \text{Re}(E_x H_y^*) dx$, is a conserved quantity in the sense that it remains constant during the propagation of surface solitons. Figure 2(a) and 2(b) plot soliton power versus propagation constant $\beta$. Notably, surface PLSs feature some critical power $P_c$, below which no solutions can be found. This property puts a minimum demand on the laser power to excite such surface solitons, which is in sharp contrast to the PLSs in the homogenous plasmonic lattice [10], as the existence of the latter requires no threshold power (i.e., $P_c$=0). As one expects, the threshold power substantially decreases with the increasing of distance of the peak-centered waveguide from the lattice boundary, as the comparisons shown in Fig. 2(a) and 2(b) for mode m=1 and m=2. The influence of metallic width on $P_c$ is also studied [Fig. 2(c)]. The dramatic increase of $P_c$ with decreasing metallic width is attributed to the strong coupling of SPPs at the opposite surfaces of each metallic layers, thus the enhanced diffraction, which in turns requires stronger nonlinearity to balance it.

We emphasize that, essentially different from the surface solitons in the truncated pure-dielectric lattice, surface PLSs reported in this Letter are not diffraction limited and thus their modal extensions could be nanoscale. To quantitatively characterize the degree of modal localization, we plot the fraction of power concentrating in the first waveguide (for m=1 modes) versus the nonlinear index change. This result is shown in Fig. 2(d). One sees that, for a moderate index change of $\Delta n$=0.05, over 90% of the soliton power are confined in the first waveguide, indicating that the transverse size of surface PLSs is less than $0.1\lambda$! Importantly, in the perfect plasmonic lattice, the same amount of index change is found to concentrate mode size being 0.17 $\lambda$ [See Fig. 1(a) and 1(b) for the profiles of solitons in homogenous plasmonic lattices], thus the energy concentration is almost double enhanced in the presence of lattice boundary. Note that such significant modal compression cannot be achieved by truncating the dielectric lattices, as there the surface lattice solitons are limited by diffraction and thus cannot get further compression if solitons are already at the subwavelength scale. Therefore, the possibility of further significant modal compression by lattice truncation is a unique property of plasmonic lattice, provided that the associated surfaces PLSs exist.

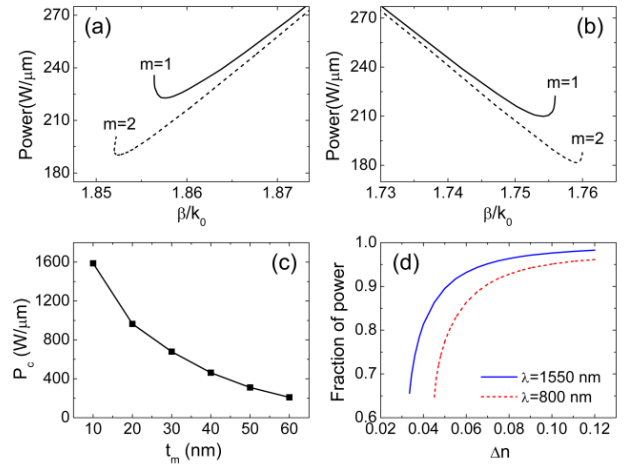

Fig. 2. (a), (b) Power vs. normalized propagation constants for staggered and unstaggered surface PLSs, respectively. (c) Threshold power $P_c$ of the unstaggered surface PLSs (m=1) as a function of metal layer width $t_m$. (d) Fraction of power concentrating in the first waveguide vs. nonlinear index change $\Delta n$, for the unstaggered surface PLSs (m=1).

To prove that the above numerically found solutions are indeed stationary eigenmodes of the associated truncated plasmonic lattice, we launch the solutions into the structure and propagate them using finite-element-method (FEM) software (Comsol Multiphysics). Typical propagation results are shown in Fig. 3(a) and 3(b); indeed the profiles remain unchanged after a long distance (> 100 μm). Moreover, for propagation results shown in Fig. 3(c) and (d), the realistic metallic loss is taken into account. From the loss-included mode analysis, for λ=1550 nm, surface PLSs have the typical absorption coefficients of ~400 cm-1 for staggered solitons, and ~250 cm-1 for unstaggered solitons which corresponds to the decay length ~25 μm and ~40 μm, respectively. As clearly seen from Fig. 3(c) and 3(d), even in the lossy case, the soliton propagation is still visible in tens of micrometers.

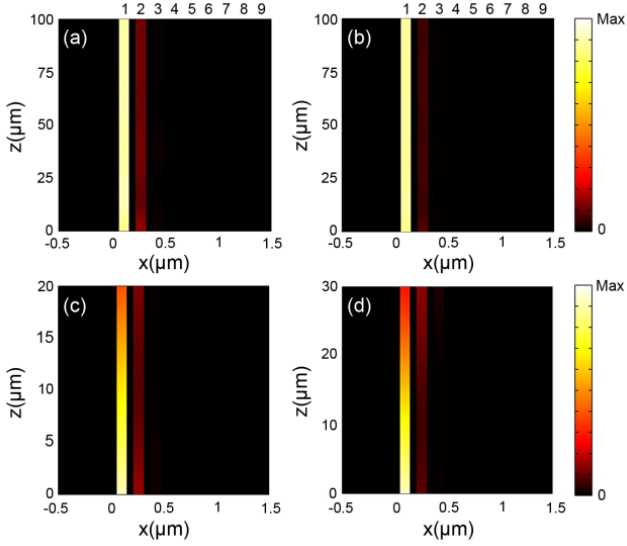

Fig. 3. Propagation of (a) staggered surface PLSs, and (b) unstagered surface PLSs in the lossless plasmonic lattices, corresponding to $\Delta n = 0.05$ ($\Delta n = -0.05$) and $\lambda=1550$ nm. (c), (d) Propagation of the same surface PLSs in the real (lossy) plasmonic lattices.

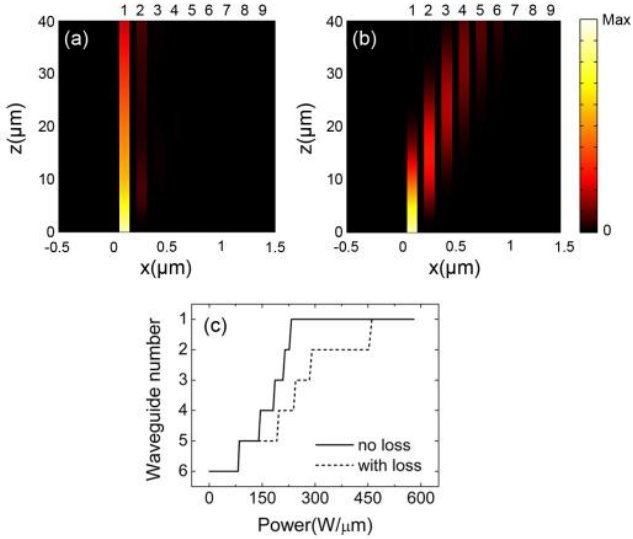

Fig. 4. Nonlinear propagation of lossy SPP beams using single-waveguide excitation with input power (a) 360 W/μm and (b) 16 W/μm. $\lambda=1550$ nm. (c) The output position of SPP beams vs. input power, for a plasmonic lattice with length of 50 μm.

We finally examine the generation of surface PLSs by means of a single-waveguide excitation setup. In Fig. 4(a) and 4(b), a TM-polarized incident light is launched into the first plasmonic waveguide. Note that, experimentally, such excitation might be achieved by, e.g., tapered couplers [15] or dipole nanoantennas [16]. We observe that, at an input power of 360 W/μm, most of light remains within the first waveguide even after a distance of 40 μm [Fig. 4(a)], indicating the formation of surface PLSs at this power level. As the input power decreases, linear diffraction gradually dominates, resulting into a significant shifting of the SPP beam towards the depth of the lattice [Fig. 4(b)]. Thus, one could precisely select the output position of light through the control of the input power. Figure 4(c) shows the dependence of the location of the peak amplitude of light at the output facet of a plasmonic lattice, in the both ideal and lossy cases. Interestingly, the figure shows that, in the lossy waveguide, switching the output position from the waveguide 6 to the waveguide 1 requires an incident power of 460 W/μm ($\Delta n \approx 0.11$), which is nearly twice the value of that in the lossless case. We finally mention that excitations from other waveguide (instead of the first waveguide) results in similar dependences of output position on input power.

In conclusion, we have studied the existence and properties as well as propagations of a new type of surface spatial solitons in a nonlinear semi-infinite plasmonic lattice, i.e., surface plasmonic lattice solitons. The impact of the lattice boundary on such surface states is elucidated. The boundary imposes a minimum power for the occurrence of surface plasmonic lattice solitons, and the truncated lattice geometry leads to a significant modal compression.

This research was supported by the National Natural Science Foundation of China (Contract No. 10874119 and 61125503) and the Foundation for Development of Science and Technology of Shanghai (Grant No. 10JC1407200 and 11XD1402600).